\documentclass[qtds]{birkmult}

\usepackage{graphicx}
\usepackage{amscd}
\usepackage{amstext}
\usepackage{amsmath}

\usepackage{amssymb,latexsym}
\usepackage{epsfig}

\newtheorem{theorem}{Theorem}[section]
\newtheorem{prop}[theorem]{Proposition}
\newtheorem{con}[theorem]{Conjecture}

\renewcommand{\theequation}{\thesection.\arabic{equation}}

\begin{document}

\title[Interaction of sources and vortices]{Interaction of point sources and vortices for incompressible planar fluids}

\author[E. A. Lacomba]{Ernesto A. Lacomba}
 \address{Department of Mathematics, UAM-I \\
  P.O.Box 55-534\\
   M\'{e}xico, D.F. 09340\\
    M\'exico}
\email{lace@xanum.uam.mx }

 \begin{abstract}
 In this paper we define the equations of motion for $N$ point
 sources as a locally Hamiltonian vector field. We compare with the
 equations for $N$ point vortices, studying first integrals and the
 blow up of collisions. Since it is a central vector field, it turns
 out that the angular momentum is identically zero; we also state
 our good conjecture that there exists a first integral in terms of
 the polar angles of the relative positions of the sources, which
 would imply that the system is always globally Hamiltonian. Then,
 we study a more general model for interaction of sources and
 vortices occurring simultaneously, which can be written as the sum
 of a Hamiltonian plus a gradient vector field.

\end{abstract}


\maketitle 




\section{Introduction}
We consider a new system of differential equations which is at the
same time gradient and locally Hamiltonian. It is obtained by just
replacing the factor $i$ by $i^2 = -1$ in the equations of
interaction for $N$ point vortices, and it is interpreted as an
interaction of $N$ point sources. Because of the local Hamiltonian
structure and the symmetries it obeys, it does possess some of the
first integrals that appear in the $N$ vortex problem. We will show
that binary collisions are easily blown up in this case since the
equations of motion are of first order. We just pass to relative
coordinates and then a convenient change of time scale. This method
may be easily generalized to the blow up of higher order collisions.

The author is indebted to Jesus Muci\~{n}o for conjecturing that a
sort of harmonic conjugate of the Hamiltonian function for point
vortices could be a first integral of the vortex system. It turns
out that this is not true, but nevertheless motivated the
introduction of the model for a system of point sources.
Furthermore, M. Hampton \cite{h} proposed to take complex
vorticities instead of the standard real ones, which allows us to
generalize the model to cover simultaneous interactions of sources
and vortices.

\section{Statement of the problem and main properties}
 The
 Hamiltonian for a system of planar point vortices is given by
\begin{equation}\label{funHam}
H = -\sum _{k=1, l > k}^N \Gamma_k \Gamma _l \ln |z_k - z_l|.
\end{equation}
If we construct the gradient vector field corresponding to $H$, we
get instead of the differential equations for $N$ point vortices (as
in Aref \cite{a} and Newton \cite{new}).
$$\dot{z}_k = i \sum _{ l \neq k}^N \Gamma _l \frac{ z _k - z _l }{ |z
_k - z _l | ^2 } = i \sum _{ l \neq k}^N \frac{\Gamma_l }{\bar{z}_k
- \bar{z}_l},$$ those corresponding to the interaction of a system
of planar point sources, i.e.
\begin{equation} \label{siste}
\Gamma_k\dot{z}_k = - \sum _{ l \neq k}^N \Gamma_k\Gamma _l \frac{ z
_k - z _l }{ |z _k - z _l | ^2 } = - \sum _{ l \neq k}^N
\frac{\Gamma_k\Gamma_l }{\bar{z}_k - \bar{z}_l}.
\end{equation}
This velocity vector field $X_s$ can also be represented as a
locally Hamiltonian vector field for the one--form
\begin{equation} \label{unoforma}
\alpha = - \textrm{d}\left(\sum _{ l > k}^N \Gamma_k\Gamma _l
\arg(z_k - z_l)\right),
\end{equation}
i.e., $i_{X_s}\omega = -\alpha,$ where $\omega =
\sum_{i=1}^{N}\Gamma_i{dx_i \wedge dy_i}.$

This $\alpha$ is globally defined in $\mathbb{C}^N
\smallsetminus\Delta$ where $\Delta = \{(z_1, \cdots, z_N)| z_j \neq
z_k \forall j, k\} $, and in fact it is a closed but not exact
one--form, since the function in between the parenthesis is only
locally defined. This is because we can rewrite
\begin{equation} \label{}
\alpha = - \sum _{ l > k}^N \Gamma_k\Gamma _l \;\textrm{d}(\arg(z_k
- z_l)),
\end{equation}
and each polar angle $\theta_{kl} = \arg(z_k - z_l)$ is defined only
modulo $2\pi$ in $\mathbb{C}\smallsetminus\{0\}$, although its
exterior differential $\textrm{d}\theta_{kl}$ is globally defined
there.

One can easily verify that the linear momentum
\begin{equation} \label{}
Z = \sum_k \Gamma_k z_k,
\end{equation}
is a first integral of system~(\ref{siste}), so if $\sum_k \Gamma_k
\neq 0$ we can define an equivalent center for the sources by $z_0 =
Z/\sum_k \Gamma_k$. On the other hand, the angular momentum $A =
\sum_k \Gamma_k z_k \times \dot{z}_k$ is an invariant of the system.
In fact, we have
\begin{prop}
The angular momentum $A$ for the point sources system is identically
zero.
\end{prop}
\begin{proof}
If $\cdot$ denotes the standard scalar product of $\mathbb{R}^2$
identified with $\mathbb{C}$, we can write
$$\sum_k \Gamma_k z_k \times \dot{z}_k = \sum_k \Gamma_k (i z_k) \cdot \dot{z}_k
= - \sum_k \Gamma_k (i z_k)\cdot \left(\sum_{l\neq k}\Gamma_l
\frac{z_k - z_l}{|z_k - z_l|^2}\right) \equiv 0,$$ since
$\Gamma_k\Gamma_l (i z_k) \cdot (z_k - z_l) + \Gamma_k\Gamma_l (i
z_l) \cdot (z_l - z_k) = 0$, for any $k\neq l$.
\end{proof}
In some sense, the reason for this result is that we are dealing
with a central velocity vector field. So, there is no simple way to
introduce rotations into the system.
\begin{con}
Along any fixed solution of~(\ref{siste}), the polar angles
$\theta_{kl}$ undergo variations smaller than $2\pi$, so that $g = -
\sum _{ l > k}^N \Gamma_k\Gamma _l\theta_{kl}$ is a first integral
of the system along the given solution.
\end{con}
However, the moment of inertia $I = \sum_k \Gamma_k |z_k|^2 = \sum_k
\Gamma_k z_k\bar{z}_k$, which is a first integral in the case of
$N$--vortices, is not always a first integral for $N$--point
sources. Indeed,
\begin{eqnarray*}
\dot{I} &=& \sum_k \Gamma_k z_k\dot{\bar{z}}_k + \sum_k \Gamma_k
\dot{z}_k\bar{z}_k \\ &=&-\sum_k z_k \sum_{l\neq k} \Gamma_k\Gamma_l
\frac{\bar{z}_k - \bar{z}_l}{|z_k - z_l|^2} -\sum_k \bar{z}_k
\sum_{l\neq k} \Gamma_k\Gamma_l \frac{z_k - z_l}{|z_k - z_l|^2}\\
&=& - \sum_{l\neq k} \Gamma_k\Gamma_l,
\end{eqnarray*} since
$z_k(\bar{z}_k - \bar{z}_l) + z_l(\bar{z}_l - \bar{z}_k) = |z_k -
z_l|^2$. So, $\dot{I}$ is in general a nonzero constant. Hence, only
the two point source problem is always integrable since it has
enough first integrals. In order for the integrability to make
sense, we have to definitely know that we are dealing with a
globally Hamiltonian vector field. But this is the case for $N = 2$,
since the angle $\theta_{12}$ is not only bounded but constant, as
it will become also clear from the analysis in the following
section.
\begin{prop}
If $\sum_{l<k} \Gamma_k\Gamma_l = 0$, then $I$ is a first integral.
If $\sum_{l<k} \Gamma_k\Gamma_l\neq 0$, then all solutions are
gradient--like with respect to $I$.
\end{prop}
This implies that only when $\sum_{l<k} \Gamma_k\Gamma_l = 0$ is
there any chance of having periodic solutions.

Comparing with the $N$--vortex problem, we recall that in that case
if the $\Gamma_k$ denote vorticities, the angular momentum takes
exactly the value $\sum_{l<k} \Gamma_k\Gamma_l$, which is called the
virial, while $I$ is a first integral of the system.

Also, collisions are rather scarce in the $N$--vortex problem and
depend on the values of the vorticities. In contrast, collisions are
very likely to occur in positive or in negative time for a system of
$N$ sources. There is no simple way to continue solutions beyond
collisions, unless we agree that any time a collision occurs the
sources involved merge into one source whose intensity is the sum of
the colliding source intensities. As a result, the solutions will
only be continuous but not smooth when collision occurs and the
dimension of the phase space jumps correspondingly.

\section{Examples: A single point source and two point sources.}
The differential equations for a single source are defined as
\begin{equation}\label{unafu}
\dot{z}= -\frac{\Gamma}{|z|^2} z = -\frac {\Gamma}{\bar{z}},
\end{equation}
where $z\in \mathbb{C}\smallsetminus\{0\}$, or in terms of real and
imaginary parts
$$\dot{x}=- \frac{\Gamma x}{x^2 + y^2}, \;\; \dot{y}= - \frac{\Gamma y}{x^2 +
y^2},$$ where $-\Gamma$ is the intensity of the point source. When
$-\Gamma < 0$ the vector field points radially towards the origin,
so that it is actually a sink; when $-\Gamma > 0$ it is really a
source, since the fluid runs away from the origin.

In this case for the local Hamiltonian formulation we can take
$\omega = \textrm{d}x \wedge \textrm{d}y$, $H = -\Gamma\ln |z|$ and
$\alpha = - \Gamma \frac{-y\textrm{d}x + x\textrm{d}y}{x^2 + y^2}$.
The one form $\alpha$ is closed but not exact on
$\mathbb{C}\smallsetminus\{0\}$, since we can write $\alpha =
-\Gamma \textrm{d}\theta$, but only when the polar angle $\theta$ is
defined in an open interval of length smaller or equal than $2\pi$.

The equations of 2 point sources of nonzero real intensities
$\Gamma_1$ and $\Gamma_2$ are given by
\begin{equation}\label{dosfu}
\dot{z}_1= -\Gamma_2\frac{z_1 - z_2}{|z_1 - z_2|^2}, \;\; \dot{z}_2=
-\Gamma_1\frac{z_2 - z_1}{|z_2 - z_1|^2}.
\end{equation}
In terms of relative coordinates $z = z_1 - z_2$, they reduce to the
single source equations of intensity $-(\Gamma_1 + \Gamma_2)$:
\begin{equation}\label{fuequi}
\dot{z}= -(\Gamma_1 + \Gamma_2)\frac{z}{|z|^2}.
\end{equation}
 If $-(\Gamma_1 +
\Gamma_2)<0$ this is an equivalent sink, so that the 2 point sources
approach each other. On the contrary, if $-(\Gamma_1 + \Gamma_2)>0$
it is an equivalent source and the 2 point sources move away from
each other. Finally, if $\Gamma_1 + \Gamma_2 = 0$, then $z$ is
constant, so that the two sources will move with the same velocity
and keeping a constant distance, as we can see from~(\ref{dosfu}).

In the cases where $\Gamma_1 + \Gamma_2 \neq 0$ we verify that
indeed $I$ is not a first integral. All the solutions will then
either end in collision or begin at collision in finite time.
Indeed, equation~(\ref{fuequi}) may be written as
$$\frac{\textrm{d}}{\textrm{d}t}|z|^2 = \dot{z}\bar{z} +
\dot{\bar{z}}z = -2 (\Gamma_1 + \Gamma_2),$$ which implies $|z|^2 =
- 2 (\Gamma_1 + \Gamma_2)t$.
\section{Regularization and blow up of binary collisions of point sources}
In this section we show how to desingularize any binary collision of
point sources.

We begin with the single point source~(\ref{unafu}), which as we saw
in the last section is equivalent to a two point source. Upon
multiplication of the equation by $\bar{z}$, we get $\bar{z}\dot{z}
= - \Gamma$, suggesting the change of time scale $\frac{dt}{d\tau} =
\bar{z}$, producing the new equation
$$\frac{dz}{d\tau} = -\Gamma$$
in terms of the time $\tau$. Moreover, we may assume $z\in
\mathbb{R}$ or $z = \bar{z}$, since any motion is radial. This
equation can be integrated as $z(\tau) = -\Gamma (\tau - \tau_0) +
z_0$, and from $dt = z(\tau) d\tau$ we get the quadratic equation
relating both times $t = -\Gamma\frac{\tau^2}{2} + (\Gamma \tau_0 +
z_0)\tau = -\frac{1}{2}\Gamma(\tau-\tau_0-z_0/\Gamma)^2 +
\frac{1}{2}\Gamma(\tau_0 + z_0/\Gamma)^2$. This may be considered as
a \emph{regularization} since the collision is not a singularity any
more and it is reached in finite time. However, the physical time is
bounded and it is not defined beyond the collision time value $t =
\frac{1}{2}\Gamma(\tau_0 + z_0/\Gamma)^2$, which is reached when
$\tau = \tau_0 + z_0/\Gamma$. This is clear, since in contrast with
the gravitational $2$--body problem, we would not expect here the
motion to be defined beyond a collapse of two sources since they
merge into a single source.

If we multiply instead equation~(\ref{unafu}) by $|z|^2$ we get a
new time rescaling $\frac{dt}{ds} = |z|^2$ which produces the
differential equation
$$\frac{dz}{ds} = -\Gamma z,$$
whose integration gives an exponential solution $z(s) = e^{-\Gamma
s}$. This new desingularization corresponds to a \emph{blow up},
since the singularity $z = 0$ is now an equilibrium point and so it
is reached in infinite time.

Because for three or more sources collision does not occur in
general along a straight line, it is the blow up which can be used
for desingularizing binary collisions. We will illustrate the
transformation for the case of 3 sources.

The differential equations for 3 sources located at $z_1, z_2,
z_3\in \mathbb{C}$ with respective intensities $\Gamma_1, \Gamma_2,
\Gamma_3$ are given by
\begin{eqnarray}\label{tresfu}
\dot{z}_1= -\Gamma_2\frac{1}{\bar{z}_1 - \bar{z}_2}
-\Gamma_3\frac{1}{\bar{z}_1 - \bar{z}_3},\nonumber\\ \dot{z}_2=
-\Gamma_1\frac{1}{\bar{z}_2 - \bar{z}_1} -\Gamma_3\frac{1}{\bar{z}_2
- \bar{z}_3},\\ \dot{z}_3= -\Gamma_1\frac{1}{\bar{z}_3 - \bar{z}_1}
-\Gamma_2\frac{1}{\bar{z}_3 - \bar{z}_2}\nonumber.
\end{eqnarray}
Passing to the relative coordinates $\xi = z_3 -  z_1, \eta = z_3 -
z_2$, we can write $z_2 - z_1 = \xi - \eta$, getting the following
system of differential equations for $\xi, \eta$
\begin{eqnarray*}
\dot{\xi} = -\frac{\Gamma_1 + \Gamma_3}{\bar{\xi}} -
\frac{\Gamma_2}{\bar{\eta}} -
\frac{\Gamma_2}{\bar{\xi}-\bar{\eta}},\\
\dot{\eta} = -\frac{\Gamma_1}{\bar{\xi}} - \frac{\Gamma_2 +
\Gamma_3}{\bar{\eta}} - \frac{\Gamma_1}{\bar{\eta} - \bar{\xi}}.
\end{eqnarray*}
The change of time scale $\frac{dt}{ds} = |\xi|^2$ yields the blow
up system
\begin{eqnarray}\label{fureg}
\frac{d\xi}{d s} = -(\Gamma_1 + \Gamma_3)\xi - \frac{\Gamma_2
|\xi|^2}{\bar{\eta}} -
\frac{\Gamma_2 |\xi|^2}{\bar{\xi}-\bar{\eta}},\\
\frac{d\eta}{d s} = -\Gamma_1 \xi - \frac{(\Gamma_2 +
\Gamma_3)|\xi|^2}{\bar{\eta}} - \frac{\Gamma_1 |\xi|^2}{\bar{\eta} -
\bar{\xi}},\nonumber
\end{eqnarray}
which is already regular at binary collision $\xi = 0$. If we want
the solution in terms of the original coordinates when $\Gamma =
\Gamma_1 + \Gamma_2 + \Gamma_3$ is nonzero we need to solve the
system
\begin{eqnarray*}
\Gamma_1 z_1+ \Gamma_2 z_2 + \Gamma_3 z_3 = \Gamma z_0,\\
-z_1 + z_3 = \xi,\\
-z_2 + z_3 = \eta,
\end{eqnarray*}
where $z_0$ is the equivalent source center for the system, getting
\begin{eqnarray*}
z_1 = z_0 + (\xi + \eta)/\Gamma - \xi,\\
z_2 = z_0 + (\xi + \eta)/\Gamma - \eta.\\
z_3 = z_0 + (\xi + \eta)/\Gamma.\\
\end{eqnarray*} If we want to blow up at the same time binary collisions
$\xi = 0$ and $\eta = 0$, we change the rescaling by $\frac{dt}{ds}
= |\xi|^2 |\eta|^2$. This situation is generalized as follows.
\begin{prop}
For binary collision blow up of $N$ sources we change to $N-1$
relative coordinates $\xi_1, \xi_2, \cdots, \xi_{N-1}$ and take a
time scale factor $\frac{dt}{ds}$ which is the product of each of
the $|\xi_i|^2$ corresponding to the collisions $\xi_i = 0$ we want
to desingularize. \end{prop}

Passing to relative coordinates is equivalent to factoring out the
translation symmetry corresponding to the first integral $Z$. One
can show that for $N$ sources, only $N-1$ suitable chosen relative
coordinates generate all the remaining ones.

If we try to formally apply this regularization and blow up
transformations to the differential equation $\dot{z}= i\frac
{\Gamma}{\bar{z}}$ for the velocity vector field of one vortex, we
get in the first case $\bar{z}\dot{z} = i\Gamma$, but rescaling does
not have any sense here if we want to keep a real time; but by
taking real parts this equation yields $\frac{d}{dt} |z|^2 = 0$. In
the case of blow up we get $\frac{dz}{ds} = |z|^2 \dot{z} = i\Gamma
z$, which upon integration gives $z(s) = K e^{i\Gamma s}$ and
finally $\frac{dt}{ds} = |z|^2 = K^2$ and so $t = K^2 s  + t_0$ . In
both cases, we see that solutions are concentric circles as it is
well known.
\subsection{Example with 3 point sources} Consider a 3--source
problem with sources initially located at the points $z_1 = 0, z_2 =
2, z_3 = i$ with corresponding intensities $\Gamma_1 = \Gamma_2 = 1$
and $\Gamma_1 = -\frac{1}{2}$. Since $Z = z_1 + z_2 -\frac{1}{2}z_3$
and $\Gamma_1 + \Gamma_2 + \Gamma_3 = \frac{3}{2}$, we get the fixed
source center $z_c = \frac{2}{3}(z_1 + z_2 -\frac{1}{2}z_3) =
\frac{4}{3} - \frac{1}{3} i$ and we have $\Gamma_1\Gamma_2 +
\Gamma_1\Gamma_3 + \Gamma_2\Gamma_3 = 0$ in this case, so that the
moment of inertia $I$ is also a first integral. Now since we have a
rectangular triangle of sides $1, 2$ and $\sqrt{5}$ which are the
mutual distances, one verifies that the corresponding initial
velocities are $v_1 = \frac{1}{2} - i, \; v_2 = -\frac{1}{10}(3 +
i), \; v_3 = \frac{2}{5}(1 - 3i)$. Notice that $v_2$ and $v_3$ are
perpendicular complex numbers and one verifies that indeed the
angular momentum is $A = 0$. However, it is easy to see that the
velocities at each of the 3 sources induce an initial
counterclockwise torque about $z_c$, which is easy to verify for
sources $1$ and $3$. On the other hand, the parametric equation of
the line through $z_2$ in the direction $v_2$ is $\alpha(k) = z_2 +
k v_2 = 2 - \frac{3}{10}k - \frac{k}{10}i$ with $k\in \mathbb{R}$.
The value $k = k_0$ such that $\textrm{Re}(\alpha(k_0)) =
\frac{4}{3}$ must satisfy $2 - \frac{3}{10}k = \frac{4}{3}$, that
is, $k_0 = \frac{20}{9}$. But then $\textrm{Im}(\alpha(k_0)) =
-\frac{k_0}{10} = -\frac{2}{9} > -\frac{1}{3}$. So that $\alpha(k)$
passes to the left of $z_2$ above the point $z_c$, as required.
Hence, the whole initial motion will be a counterclockwise rotation
about $z_c$. A numerical simulation of the dynamics using
Mathematica shows that a binary collision of sources $1$ and $3$
occurs at some positive time, merging into a sink of intensity
$\frac{1}{2}$. This happens without any significant rotation of the
source system. We can apply a blow up for this collision, obtaining
equations~(\ref{fureg}) in the new time $s$. Simulation with this
blow up system gives already approximately the asymptotic positions
of the sources for $s = 15$ or $s = 30$ as $z_1(30) = z_3(30) =  0.5
- 0.773 i$ and $z_2(30) = 1.75 - 0.1133 i$.
\section{Interaction of point sources and point vortices} We
can generalize the models for interaction of point vortices and of
point sources to a model where we consider simultaneous interaction
of point vortices and point sources.

The idea is to formally write the differential equations for point
vortices where we assume now that the vorticities are nonzero
complex numbers, that is
$$\dot{z}_k = i \sum _{ l \neq k}^N \Gamma _l \frac{ z _k - z _l }{ |z
_k - z _l | ^2 } ,$$ with $\Gamma_k \in\mathbb{C}\smallsetminus{0}$.
This new velocity vector field $X$ can be written as the sum of a
Hamiltonian vector field $X_1$ with respect to the function $$H =
-\sum _{k=1, l
> k}^N \textrm{Re}(\Gamma_k) \textrm{Re}(\Gamma _l)\ln |z_k -
z_l|,$$ plus a gradient vector field $X_2$ with respect to the
function $$G = -\sum _{k=1, l
> k}^N \textrm{Im}(\Gamma_k) \textrm{Im}(\Gamma _l)\ln |z_k -
z_l|,$$ if we decompose the system as
$$\dot{z}_k = i \sum _{ l \neq k}^N \textrm{Re}(\Gamma_k)
\frac{ z _k - z _l }{ |z _k - z _l | ^2 } - \sum _{ l \neq k}^N
\textrm{Im}(\Gamma _l) \frac{ z _k - z _l }{ |z _k - z _l | ^2 }.$$
The vector field $X_2$ is also locally Hamiltonian with respect to
the one--form $\alpha = - \sum _{ l > k}^N
\textrm{Im}(\Gamma_k)\textrm{Im}(\Gamma_l) \;\textrm{d}(\arg(z_k -
z_l)),$ globally defined in $\mathbb{C}^N \smallsetminus\Delta$.
Here, $\textrm{Re}(\Gamma_k)$ represent the vorticities and
$-\textrm{Im}(\Gamma_k)$ represent the source intensities. This is a
particular case of the Helmholtz or Stokes decomposition of volume
preserving velocity vector fields as the sum of a rotational plus a
gradient term.

If not all the real numbers $\textrm{Re}(\Gamma_k)$ are nonzero, we
reduce correspondingly the dimension of phase space in order to
define the corresponding symplectic structure to produce the
Hamiltonian vector field $X_1$. Similarly for the independent
symplectic structure making $X_2$ locally Hamiltonian when some
$\textrm{Im}(\Gamma_k)$ are zero.
\subsection{Example, 2 point source-vortices}
The equations of 2 point source-vortices of nonzero complex
intensities $\Gamma_1$ and $\Gamma_2$ are given by
\begin{equation}\label{dosfuvo}
\dot{z}_1= -\Gamma_2\frac{z_1 - z_2}{|z_1 - z_2|^2}, \;\; \dot{z}_2=
-\Gamma_1\frac{z_2 - z_1}{|z_2 - z_1|^2}.
\end{equation}
In terms of relative coordinates $z = z_1 - z_2$, they reduce to the
single source-vortex equations of intensity $-(\Gamma_1 +
\Gamma_2)$:
\begin{equation}\label{fuvoequi}
\dot{z}= -(\Gamma_1 + \Gamma_2)\frac{z}{|z|^2}.
\end{equation}
The velocity vector field corresponding to these equations is
considered in Chapter 11 of Needham \cite{nee} in a different
context. This time equation~(\ref{fuvoequi}) implies $|z|^2 = - 2
\textrm{Re}(\Gamma_1 + \Gamma_2)t + r_0^2.$ But in fact, if we write
$z = r e^{i\theta}$ then $\dot{z} = (\dot{r} + i r \dot{\theta})
e^{i\theta}$ or $\bar{z}\dot{z} = (r\dot{r} + i r^2 \dot{\theta}) =
-(\Gamma_1 + \Gamma_2),$ i.e. $r\dot{r} = - \textrm{Re}(\Gamma_1 +
\Gamma_2), r^2 \dot{\theta} = - \textrm{Im}(\Gamma_1 + \Gamma_2).$
Integrating, we get
$$ r = \sqrt{-2 t\textrm{Re}(\Gamma_1 + \Gamma_2) + r_0^2}$$ and
$$\theta = \int \frac{-\textrm{Im}(\Gamma_1 + \Gamma_2) dt}{-2 t\textrm{Re}(\Gamma_1 + \Gamma_2) + r_0^2} + \theta_0.$$
This corresponds to a logarithmic spiral of the form $r = K
e^{\alpha\theta}$. So that a collision occurs in finite time with
infinite spiralling, that is, while the argument $\theta$ remains
unbounded. If $\textrm{Re}(\Gamma_1 + \Gamma_2) = 0,$ then $r = r_0$
is constant and $\theta$ is linear in $t,$ corresponding to an
equivalent pure vortex. Likewise, if $\textrm{Im}(\Gamma_1 +
\Gamma_2) = 0,$ then $\theta = \theta_0$ is constant and $r^2$ is
linear in $t,$ which is equivalent to a pure source.

If we apply now a blow up to equation~(\ref{fuvoequi}), we get
$|z|^2 \dot{z} = -(\Gamma_1 + \Gamma_2) z$. The change of time scale
$\frac{dt}{ds} = |z|^2$ gives $\frac{dz}{ds} = -(\Gamma_1 +
\Gamma_2) z$, which is integrated as $z(s) = K e^{-(\Gamma_1 +
\Gamma_2) s} = K e^{-\textrm{Re}(\Gamma_1 + \Gamma_2) s}
e^{-i\textrm{Im}(\Gamma_1 + \Gamma_2) s}$, giving another
description of the logarithmic spiral in terms of time $s$. We
remark that an infinite spiraling while going to collision in finite
time  occurs also in the Manev problem which is an approximate model
for the Schwarschild equation in general relativity.

\subsection*{Acknowledgements} This research has been partially
supported by CONACYT-M\'exico, grant 47768 and by a PIFI 2007
project UAM-I-CA-55 ``Differential Equations and Geometry".

\end{document}